\documentstyle[12pt]{article}
\begin{document}
\newcommand{\newc}{\newcommand}
\newc{\ux}{$U(1)_X$\ }
\newc{\mpl}{$M_{Pl}$\ }
\newc{\mx}{$M_X$\ }
\newcommand{\del}{\partial}
\newc{\beq}{\begin{equation}}
\newc{\eeq}{\end{equation}}
\newc{\barr}{\begin{eqnarray}}
\newc{\earr}{\end{eqnarray}}
\newc{\ra}{\rightarrow}
\newc{\lam}{\lambda}
\newc{\eps}{\epsilon}
\newc{\phot}{{\tilde\gamma}}
\newc{\half}{\frac{1}{2}}
\newc{\third}{\frac{1}{3}}
\newc{\fourth}{\frac{1}{4}}
\newc{\eighth}{\frac{1}{8}}
\newc{\gev}{\,GeV}
\newc{\tev}{\,TeV}
\newc{\lra}{\leftrightarrow}
\newc{\Dslash}{\not\!\! D}
\newc{\sg}{{\cal G}}
\newc{\eq}[1]{(\ref{eq:#1})}
\newc{\eqs}[2]{(\ref{eq:#1},\ref{eq:#2})}
\newc{\lab}[1]{\label{eq:#1}}
\newc{\sect}[1]{\ref{sec:#1}}
\newc{\etal}{{\it et al.}\ }
\newc{\eg}{{\it e.g.}\ }
\newc{\ie}{{\it i.e.}\ }
\newc{\nonum}{\nonumber}
\newc{\kap}{\kappa}
\newc{\Hbar}{{\bar H}}
\newc{\hhbar}{{\overline h}}
\newc{\Ubar}{{\bar U}}
\newc{\Dbar}{{\bar D}}
\newc{\Ebar}{{\bar E}}
\newc{\vev}[1]{<\!{#1}\!>}

\title{Anomaly-Free Gauged $U(1)'$ in Local Supersymmetry and
Baryon-Number Violation}
\author{A. H. Chamseddine and Herbi Dreiner}
\date{{\small Theoretische Physik, ETH-H\"onggerberg, CH-8093 Z\"urich,
Switzerland}}
\maketitle

\vspace{-6.5cm}
\hfill\parbox{8cm}{\raggedleft March, 28$^{th}$, 1995 \\ ETH-TH/95-06}
\vspace{6.5cm}

\begin{abstract}
\noindent
The supersymmetric extension of the standard model suffers from a problem
of baryon-number violation. Discrete (and global) symmetries introduced to
protect the proton are unstable under gravitational effects. We add a gauged
$U(1)_X$ to the standard model gauge group $G_{SM}$ and require it to be
anomaly-free. As new (chiral) superfields we only allow $G_{SM}$-singlets
in order to maintain the good unification predictions. We find the most
general set of solutions for the rational singlet charges. We embed our
models in {\it local} supersymmetry and study the breaking of supersymmetry
and $U(1)_X$ to determine $M_X$. We determine the full non-renormalizable
and gauge invariant Lagrangian for the different solutions. We expect any
effective theory to contain baryon- and lepton-number violating terms of
dimension four suppressed by powers of $M_X/M_{Pl}$. The power is predicted
by the $U(1)_X$ charges. We find consistency with the experimental bounds on
the proton lifetime and on the neutrino masses. We also expect all
supersymmetric models to have an unstable but longlived lightest supersymmetric
particle. Consistency with underground experiments on upward going muons leads
to stricter constraints than the proton decay experiments. These are barely
satisfied.
\end{abstract}

\section{Introduction}
When incorporating supersymmetry into the Standard Model one immediately
runs into a problem. The Standard Model conserves baryon- ($B$) and
lepton-number ($L$) automatically and higher dimensional $\Delta B,\Delta
L\not=0$ operators are suppressed by at least 4 powers of the scale of
baryon- or lepton-number violation. However, in supersymmetry
\cite{nilles} the most general interactions involving the Standard Model
(super-) fields and invariant under the Standard Model gauge group
\beq
G_{SM}=SU(3)_C\otimes SU(2)_L\otimes U(1)_Y,
\eeq
include $\Delta B,\Delta L\not=0$ terms of dimension 4
\cite{weinberg,sy,drw}
\beq
(LL\Ebar)_F,\quad (LQ\Dbar)_F,\quad(\Ubar\Dbar\Dbar)_F,\quad(L\Hbar)_F .
\lab{dim4}
\eeq
Here $L,Q$ are the left-handed lepton and quark superfields
respectively.\footnote{We have used the same symbol $L$ for lepton number
and for the left-handed lepton doublet. The context should always make it
clear which is meant.} $\Ebar,\Ubar,\Dbar$ are the corresponding
right-handed superfields. Their $G_{SM}$ quantum numbers are given below
in Eq.\eq{qmnumbers}. There are also dim 5 terms which can lead to
dangerous levels of proton decay \cite{nilles,weinberg}
\barr
&&(LL{\Hbar}\Hbar)_F,\quad(QQQL)_F,\quad (\Ubar\Ubar\Dbar\Ebar)_F,\quad
(LH\Hbar\Hbar)_F,\nonumber \\ &&
(QQQH)_F,\quad(HQ\Ubar\Ebar)_F \quad
(Q\Ubar L^*)_D,\quad (\Ubar\Dbar^*\Ebar)_D, \lab{dim5}\\
&&(\Hbar\Hbar\Ebar^*)_D,\quad (\Hbar^*H\Ebar)_D.\nonumber
\earr

With the gauge structure restricted to that of the minimal supersymmetric
standard model (MSSM) the $L\Hbar$ term
in Eq.\eq{dim4} can be rotated away through a field redefinition. It is
then absorbed in the $LL\Ebar$ and $LQ\Dbar$ terms as well as those of the
MSSM. However, if extra gauge symmetries are present which distinguish
between $L$ and $H$ this is no longer true. This gauge symmetry could
prohibit the first two terms of Eq.\eq{dim4} as well as the $\mu H\Hbar$
term of the MSSM while allowing a $\kappa L\Hbar$ term. This can then no
longer be rotated away. In the low-energy effective theory this extra
symmetry is
broken. If it is broken at sufficiently high energy no effects are
observable and the $L\Hbar$ term can again be rotated away. We thus retain
the $L\Hbar$ term in our discussion and decide if there is a remnant
effective $L\Hbar$ term.

The dim 4-terms together lead to an unacceptable level of proton decay.
Thus the symmetry of the SM must be extended to
\beq
G=G_{SM}\otimes {\tilde G},
\eeq
such that ${\tilde G}$ guarantees the longevity of the proton. Discrete,
global, and gauge symmetries have been considered. The first solution to
the problem of proton decay was to introduce the discrete symmetry
R-parity \cite{dimgeorgi,sy,drw} which prohibits all the terms in
\eq{dim4}. Later other discrete symmetries were considered, in particular
baryon-parity \cite{ibross} which only prohibits the term $\Ubar
\Dbar\Dbar$. This is sufficient to protect the proton. However, Krauss and
Wilczek \cite{krauss} later showed that discrete symmetries are violated
by gravitational effects unless they are remnants of a broken gauge
symmetry. So only very special discrete symmetries, discrete gauge
symmteries, are viable candidates for ${\tilde G}$. Ibanez and Ross
\cite{ibanross1,ibross} translated the requirement of an anomaly-free
gauge theory into a set of conditions on the remnant discrete symmetry.
They then systematically studied all family-independent discrete
symmetries up to ${\bf Z}_4$. They found that only R-parity and
baryon-parity are anomaly-free in this sense. Baryon-parity has the
advantage of forbidding the dangerous dim-5 $\Delta B\not=0$ operators
\eq{dim5}. Both symmetries are consistent with all known experiments but
lead to drastically different predictions for future collider searches of
supersymmetry \cite{dreiner,morawitz}.

Global symmetries suffer from the same gravitational breaking as discrete
symmetries. However, they can not be the remnant of an anomaly-free broken
gauge symmetry. In addition a global baryon number for example leads to
problems with baryogenesis and if broken a global symmetry can lead
to an unwanted axion. Therefore we do not further consider global
symmetries.

Thus it is most likely that the proton is protected by a (broken) gauge
symmetry. A first solution to the problem of baryon-number violation via a
gauge symmetry ${\tilde G}=U(1)_X$ was considered by S. Weinberg
\cite{weinberg}. His approach was to give all the matter fields charges
$Q_X$ of the same sign and the Higgs fields the opposite sign.\footnote{This
is just the gauged Fayet $U(1)$ \cite{fayet} which was introduced in order
to give the scalar fermions large positive mass corrections after supersymmetry
breaking.} In order to guarantee an anomaly-free theory it is then necessary to
include fields which transform non-trvially under $G_{SM}$, in particular
also additional ($SU(3)$) coloured fields. Similar models have later
been constructed by \cite{japan}. However, the extra coloured field must be
massive and such a mass term is in general not $U(1)_X$ gauge invariant.
In principle it can however get a heavy mass when $U(1)_X$ is broken.

Besides the problem of giving $G_{SM}$ non-singlets a mass such fields
also drastically affect the success of the unification of the gauge
couplings in supersymmetry provided their masses are below the
unification scale. We shall thus only consider $G_{SM}$-singlets as
additional fields to cancel the \ux\ anomalies. In this case there is no
longer an anomaly-free $U(1)_X$  where all matter fields have the same
charge and the problem of $\Delta B,\Delta L\not=0$ interactions must be
reconsidered.

This approach was first studied by Font \etal\ \cite{font}. The authors
searched for a $U(1)_X$ which prohibits all dangerous dimension four terms
\eq{dim4}. They found that \ux\  necessarily acts on the MSSM matter
fields as a linear combination of hypercharge $Y$ and the third component
of right-handed iso-spin $I_3^R$. Since $(B-L)$ is a linear combination of
$Y$ and $I_3^R$, $U(1)_X$ can also be considered as a linear combination
of $Y$ and $(B-L)$
\beq
X=\alpha_Y Y+\alpha_{(B-L)} {(B-L)}+
\alpha_{S} {S},
\lab{linu1}
\eeq
here $S$ is the part of $U(1)_X$ acting on the singlets. Later, in
\cite{ibanross1,ibross} this work was extended to search for a $U(1)_X$
which only prohibits a subclass of the terms \eq{dim4}. They found both
a gauged baryon-number and a gauged lepton-number. However, the
anomaly-free solutions require additional fields (beyond those of MSSM)
transforming non-trivially under $G_{SM}$, for example the leptoquark of
$E_6$. As stated above, we do not further consider this approach here.

Since in supersymmetry the proton is most likely protected by a (broken)
gauge symmetry, we study the possible \ux\ solutions in detail. We extend
the work of \cite{font} in several points. First, we determine the most
general $U(1)_S$ of \eq{linu1} which is anomaly-free. For three additional
$G_{SM}$ singlets we find an infinite set of rational solutions. The
singlet charges are generally not identical and thus an interpretation in
terms of right-handed neutrinos and an embedding in (a family independent)
$SO(10)$ is not necessary. Second, due to the importence of gravitational
effects when considering discrete symmetries we embed our models in local
supersymmetry. We consider local supersymmetry as an effective theory of a
more complete unified theory including gravity. We can then study the
spontaneous breaking of supersymmetry and \ux\ explicitely. We determine
the scale of \ux\ breaking (\mx) dynamically in terms of the Planck scale
\mpl and the supersymmetry scale. Before \ux\ is broken, but below \mpl
the superpotential can in principle contain {\it all non-renormalizable}
terms compatible with $G_{SM}\otimes U(1)_X$ suppressed by powers of
$M_{Pl}^{-1}$. These terms could be generated by loop effects from the
broken unified theory. We consider all such terms and their possible
effects at low energies, eventhough in any explicit (yet to be constructed
and therefore unknown) unified model we expect only a subset of these
terms to be generated via loop effects. After the breaking of \ux\ at \mx,
the renormalizable superpotential will then {\it always} contain all terms
\eq{dim4} suppressed by powers of $(M_X/M_{Pl})$. We thus predict the
scale of $B$ and $L$ violating effects including neutrino masses in the
effective theory and find them compatible with experiment. An interesting
effect of this scenario is that we always find an unstable but long-lived
lightest supersymmetric particle (LSP). We find the cosmological
constraints on the LSP lifetime to be significantly stricter than those
due to proton decay and only barely compatible with our models.

The above discussion refers to $U(1)$ gauge symmetries. In a separate
publication we discuss gauged $U(1)$ R-symmetries \cite{aliherbi}.

\section{Anomaly Cancellation Conditions}
\label{sec:cancel}
We shall embed our $G_{SM}\otimes U(1)_X$ models in $N=1$ local supersymmetry.
In accordance with our philosophy stated above, the matter chiral multiplets
are those of the MSSM with the addition of $G_{SM}$ singlets, $N,$ and $Z_m$.
These multiplets are denoted by
\beq
\begin{array}{rlrlrl}
L:& (1,2,-\frac{1}{2},l),\qquad&\Ebar: & (1,1,1,e),\qquad
     &Q:& (3,2,\frac{1}{6},q), \\ &&&&\\
\Ubar:& ({\bar3},1,-\frac{2}{3},u),\qquad&
\Dbar:& ({\bar3},1,\frac{1}{3},d), \qquad &
H:& (1,2,-\frac{1}{2},h), \\ &&&&\\
{\bar H}:& (1,{\bar2},\frac{1}{2},{\bar h}) \qquad
&N:&  (1,1,0,n),\qquad
&Z_m:&  (1,1,0,z_m),
\end{array}
\lab{qmnumbers}
\eeq
where we have indicated in parentheses the $G_{SM}\otimes U(1)_X$ quantum
numbers. We shall assume that the superpotential in the observable sector
has the form
\beq
g^{(O)}= h_{E}^{ij}L^i{E^c}^jH + h_{D}^{ij}Q^i{D^c}^jH +
h_{U}^{ij}Q^i{U^c}^j{\bar H}+ h_N NH{\bar H}
\eeq
where $h_E,\,h_D,\,h_U$ $h_N$ and are Yukawa couplings. Thus at this stage we
assume the theory conserves R-parity. We have added the term $NH{\bar H}$
instead of $\mu H{\bar H}$ as in the MSSM, in order to incorporate a possible
solution to the $\mu$-problem. We shall show below that this is not possible.
The singlet couplings will be determined by the charges $z_m$ obtained from the
solutions to the anomaly equations below.

To build a realistic model the superpotential should be gauge-invariant and
the new gauge symmetry \ux\ should be anomaly-free, \ie
\barr
l+e+h&=&0, \lab{emass}\\
q+d+h&=&0,\lab{dmass}\\
q+u+{\bar h}&=&0,\lab{umass}\\
n+h+{\bar h}&=&0,\lab{muterm}
\earr
\vspace{-1.cm}
\barr
3[\frac{1}{2}l+e+\frac{1}{6}q+\frac{4}{3}u+\frac{1}{3}d] +
\frac{1}{2}(h+{\bar h}) &=&0, \lab{yyr}\\
3[-l^2+e^2+q^2-2u^2+d^2]-h^2+{\bar h}^2 &=&0, \lab{yrr}\\
3[2l^3+e^3+6q^3+3u^3+3d^3]+ 2h^3+2{\bar h}^3
+n^3+\sum z_m^3&=&0, \lab{rrr}\\
3[\frac{1}{2}l+\frac{3}{2}q]+\frac{1}{2}(h+{\bar h})& = &0,
\lab{sulr}\\
3(q+\frac{1}{2}u+\frac{1}{2}d)&=&0,
\lab{sucr}\\
3[2l+e+6q+3u+3d]+2(h+{\bar h})+n+\sum z_m&=&0.
\lab{gravr}
\earr
The last six equations are the $Y^2X,YX^2,X^3,(SU(2)_L)^2X,(SU(3)_c)^2X$ and
gravitational anomaly \cite{gravity} equations, respectively. We have assumed
that the
\ux-charges are family independent, \eg $l_1=l_2=l_3=l$ and
there are three generations. In solving this set
of ten equations we first notice that the  Eqs.\eq{emass}-\eq{umass},
\eqs{yyr}{yrr}, \eq{sulr}, and \eq{sucr} are independent of the singlets
$N,Z_m$. The solution to these seven equation can be expressed
in terms of two variables which we choose to be $l$ and $e$
\beq
h=-l-e, \quad {\bar h}= l+e,\quad
q=-\frac{1}{3} l,\quad u=-\frac{2}{3}l-e,\quad
d=\frac{4}{3}l +e. \lab{dle}
\eeq
Inserting the values for $h$ and $\hhbar$ into Eq.\eq{muterm} we obtain
\beq
n=0.
\eeq
Since the \ux-number of the singlet $N$ is zero it does not affect the
anomalies and can be discounted. We thus eliminate it from our further
discussion and replace $NH\Hbar$ in $g^{(O)}$ by
\beq
\mu H{\bar H}.
\eeq
We thus offer no new insights on the $\mu$-problem. However, there is also
no Peccei-Quinn axion in the theory.
The remaining equations involving the singlet field charges are
\barr
3(2l+e)+\sum z_m&=&0, \lab{linuone}\\
3(2l+e)^3+\sum z_m^3&=&0.
\lab{cubicuone}
\earr
The choice $l=-e/2$ would give the chiral multiplets in the observable sector
\ux-numbers which are identical to the Y-numbers. Thus, non trivial solutions
($U(1)_X\not=U(1)_Y$) are only possible through the inclusion of the singlets
$Z_m$. We also can see from the above equations that with the {\it minimal}
field content (no singlets) charge quantization ($q_d=q_e/3$ {\it etc.}) is
given as the unique solution to the anomaly equations. This is true in the
SM and MSSM and independently of the gravitational anomaly equation
\cite{marshak,font}.

The cases of $l=-e/2$ with non-trvial singlet charges are not very interesting
and we shall require $2l+e\not=0$. For only one singlet there is no solution to
Eqs.\eqs{linuone}{cubicuone}. For two singlets there are no real solutions. The
only real solution of the cubic equation in \eq{cubicuone} has $e=-2l$. For
three singlets an obvious solution is
\beq
z_1=z_2=z_3=-(2l+e).\lab{bl}
\eeq
We shall normalize $z_1=1$ since the equations contain no constant term. For
\beq
z_2=\frac{1}{2}(x+y),\qquad z_3=\frac{1}{2}(y-x),
\eeq
we then obtain the quadratic equation in $x$:
\barr
2l+e&=&-\frac{1}{3}(y+1), \lab{ylin}\\
27yx^2+(y-2)^2(5y+8)&=&0. \lab{yquad}
\earr
There are two classes of solutions. For $y=2$ we must have $x=0$; this
corresponds to the charges
\beq
2l+e=-1,\quad z_1=z_2=z_3=1,
\lab{y2}
\eeq
This is the same as \eq{bl}. For $l=-1$ this corresponds to the gauged $(B-L)$
symmetry of $SO(10)$ and the three singlets are interpreted as the right-handed
neutrinos of the ${\bf 16}_{SO(10)}$.

The other class has $y\not=2$. The condition for a
real solution of Eq.\eq{yquad} is $y\eps\,[-\frac{8}{5},0)$. In
order to have a rational solution we must have
\beq
\frac{5y+8}{3y}=-q^2,\quad or \quad y=-\frac{8}{5+3q^2},
\qquad q\,\eps\,{\bf Q}.
\eeq
Then
\beq
x=\pm\frac{q}{3}\left(\frac{8}{5+3q^2}+2\right).
\eeq
The two signs for $x$ correspond to the interchange $z_2\leftrightarrow z_3$.
Choosing the plus sign and solving for the singlet charges we obtain
\beq
z_1=1, \quad z_2=\frac{(q-1)(q^2+q+4)}{5+3q^2},
\quad
z_3=-\frac{(q+1)(q^2-q+4)}{5+3q^2}.\lab{singlet}
\eeq
The charges for the standard model fields are then determined by two
free parameters $l$ and $q$ via
\beq
e=\frac{1-q^2}{5+3q^2}-2l.
\lab{eq}
\eeq
We have thus obtained the complete set of anomaly-free solutions for
$U(1)_X$ and three additional singlets.\footnote{The case $2l+e=0$ is included
as $q=\pm1$, $z_1=1,\,z_2=0,\,z_3=-1$.} When discussing the details of models
we shall mostly focus on the specific solutions
\barr
q=0,&&\quad 2l+e=\frac{1}{5},\quad z_1=1,z_2=z_3=-\frac{4}{5}, \lab{mod1}\\
q=3,&&\quad 2l+e=-\frac{1}{4},\quad z_1=z_2=1,z_3=-\frac{5}{4}. \lab{mod2}
\earr
The other models lead to very similar results as we shall see. We can use these
solutions to determine the coefficients $\alpha_i$ of \eq{linu1}. The
additional
$U(1)$'s which only act on the singlets are given by
\barr
U(1)_{S_1}:\quad (z_1,z_2,z_3)&=& (1,-1,0),\\
U(1)_{S_2}:\quad (z_1,z_2,z_3)&=& (1,0,-1),
\earr
and we have
\barr
U(1)_X&=&2(l+e)U(1)_Y-(2l+e)U(1)_{(B-L)}\\&+&\third (1-2z_2+z_3)U(1)_{S_1}
+ \third (z_2-2z_3+1)U(1)_{S_2}.\lab{u1ex}
\earr
This contains all solutions, including \eq{y2}. For the solutions \eq{singlet}
the $\alpha_i$ are given in terms of $l,q$
\barr
\alpha_Y&=&\left(\frac{1-q^2}{5+3q^2}-l\right)
,\quad\alpha_{(B-L)}=-\frac{1-q^2}{5+3q^2},\nonumber\\
\alpha_{S_1}&=&-\frac{q^3-q^2+3q-3}{5+3q^2},\quad
\alpha_{S_2}=\frac{q^3+q^2+3q+3}{5+3q^2}.
\earr
Thus we have found the following result. Due to the free parameter $l$ above,
and taking the $z_1=1$ normalization into account, any rational linear
combination
of $U(1)_Y$ and $U(1)_{B-L}$ can be made anomaly-free acting on the MSSM fields
with three additional $G_{SM}$ singlets. For each linear combination of
$U(1)_Y$
and  $U(1)_{B-L}$ there is an infinite set of charges $z_1,z_2,z_3$ given
by \eq{singlet} which lead to an anomaly-free $G_{SM}\otimes U(1)_X$ model.
These can be expressed as specific rational linear combination of
$U(1)_{S_1,S_2}$. Next we dynamically break supersymmetry and $U(1)_X$.

\section{Breaking of Supersymmetry and \ux }
To have a realistic model both supersymmetry and \ux must be broken at
low energies. Since we have a locally supersymmetric theory, it is possible to
break supersymmetry spontaneously. The easiest way is to utilize a hidden
sector whose fields are singlets with respect to the Standard Model gauge
group. Depending on whether the \ux-symmetry and supersymmetry are to be broken
simultaneously or not, these singlets would have or not have non-trivial
\ux-numbers.

We can break supersymmetry independently of the \ux\ by adding to the
system one singlet $Z$ with zero \ux-number. This clearly has no affect on
the anomaly equations. Then we can take the superpotential
\beq
g=m^2(Z+\beta)+g'(Z_1,Z_2,Z_3)+g^{(O)}(S_i)+g''(Z_i,S_j),
\eeq
where $g^{(O)}(S_i)$ is the observable sector superpotential which only
depends on the SM chiral superfields $S_i$. $g''(Z_i,S_j)$ is a
non-renormalizable part of the superpotential involving also interactions
between $Z_i$ and $S_j$. We will discuss it in more detail in the next
section. The first term is the Polonyi
potential, where $\beta$ is a constant. It will be fine-tuned so that the
cosmological constant is zero.

In the following we shall study the two models \eqs{mod1}{mod2} from the
previous section. The other models will give similar results as long as
two singlets have different sign charges. This is always the case for
\eq{singlet}. It is not the case for the solutions \eq{bl} which correspond
to $U(1)_X=-(2l+e)(B-L)$. In this case it is not possible to write a
superpotential of weight zero for the singlet fields.

For the model \eq{mod2} it is not possible to write a renormalizable
potential as a function of $Z_1,Z_2,Z_3$. The simplest function we can write is
\beq
g'=\kap^6Z_1^4\left( \sum_{p=0}^5 a_p Z_2^{5-p}Z_3^p,
\right) \lab{gprime}
\eeq
and we can assume the symmetry $Z_1\lra Z_2$ which implies that
$a_p=a_{5-p}$. We have inserted the Planck scale in \eq{gprime} in order
to avoid introducing new scales, $\kap=\sqrt{8\pi G_N}\approx
10^{-19}\gev^{-1}$. For the model \eq{mod1} the superpotential is given by
\beq
g'=\kap^6 Z_3^5 \sum_{p=0}^4 a_p Z_1^{4-p} Z_2^{p}.
\eeq
The two models lead to very similar results.

We now analyze the potential $V(Z,Z_1,Z_2,Z_3)$ with $g'$ given in
Eq.\eq{gprime}. We shall take the K\"ahler function to correspond to minimal
kinetic energy for the scalar fields
\beq
{\cal K}= -\frac{\kap^2}{2} (|Z|^2+|Z_1|^2+|Z_2|^2+|Z_3|^2+|Z_i|^2).
\eeq
If we ignore the low-energy fields $S_i$, \ie consider
$g'',g^{(O)}\!\!\ll\,\, \vev{Z}^3,\vev{Z_i}^3$ we obtain for the potential
\barr
V&=& \half e^{\half\kap^2|Z_a|^2}
\left[\, \left| m^2 + \frac{\kap^2}{2}{\bar Z} \left(m^2(Z+\beta) +
\kap^2Z_3^4\sum_{p=0}^5 a_p Z_1^{5-p}Z_2^p\right)\,\right|^2\right. \nonum \\
&+&\left|\kap^6Z_3^4\sum_{p=0}^4(5-p)a_pZ_1^{4-p}Z_2^p+
  \frac{\kap^2}{2} {\bar Z}_1 \left(m^2(Z+\beta)+\kap^6Z_3^4\sum_{p=0}^5
   a_pZ_1^{5-p}Z_2^p\right)\, \right|^2  \nonum \\
&+& \left|\kap^6Z_3^4\sum_{p=1}^5pa_pZ_1^{5-p}Z_2^{p-1}
  +\frac{\kap^2}{2} {\bar Z}_2 \left(m^2(Z+\beta)+\kap^6Z_3^4\sum_{p=0}^5
  a_pZ_1^{5-p}Z_2^p \right)\,\right|^2 \nonum\\
&+&  \left|4\kap^6Z_3^3\sum_{p=0}^5 a_p Z_1^{5-p}Z_2^p
  +\frac{\kap^2}{2}{\bar Z}_3\left( m^2(Z+\beta)+\kap^6 Z_3^4
  \sum_{p=0}^5 a_pZ_1^{5-p} Z_2^p\right)\,\right|^2 \nonum\\
&-&\left.  \frac{3}{2}\kap^2\left| m^2(Z+\beta)+\kap^6 Z_3^4 \sum_{p=0}^5
  a_pZ_1^{5-p}Z_2^p\right|^2\, \right] \nonum\\
&+& (\frac{{{\tilde g}}^{'2}}{8}) \left| |Z_1|^2+|Z_2|^2-\frac{5}{4}|Z_3|^2
\right|^2
\earr
${\tilde g}'$ is the \ux\ gauge coupling, ${\bar Z}$ is the complex
conjugate field of $Z$.
We have assumed here that the gauge field kinetic energy term is minimal:
$f_{\alpha\beta}=\delta_{\alpha\beta}$. Since the fields $Z_i$ transform
under the extra \ux\ group the corresponding D-term appears as the
last term in the above potential. There are many possible minima for this
potential, mainly
with
$\vev{Z_1}\approx\vev{Z_2}\approx\vev{Z_3}\approx\vev{Z}\approx\frac{1}{\kap}$.
However, we are mainly interested in the situation where
$\vev{Z}\approx\frac{1}{\kap}\gg\vev{Z_1},\vev{Z_2},\vev{Z_3}$ and we shall
tune
$\beta$ and the constants $a_p$ so that $V$ is zero at such a minimum
and positive definite, and where the condition
\beq
|Z_1|^2+|Z_2|^2 \simeq \frac{5}{4} |Z_3|^2
\eeq
is satisfied. In this case we can expand around the vev
$Z\approx\frac{1}{\kap}$
and write down the effective potantial as a function of $Z_1,Z_2$, and $Z_3$
\cite{ali}
\barr
V&=& |{\hat g}_{,i}|^2 + m_{3/2}^2 |Z_i|^2+m_{3/2}\left[Z^i{\hat g}_{,i}
+(A-3){\hat g} +h.c.\right] \nonum \\
&&+ \frac{{\tilde g}^{'2}}{8} \left|
|Z_1|^2+|Z_2|^2-\frac{5}{4}|Z_3|^2\right|^2
\lab{veffuone}
\earr
where ${\hat g}$ is the same as $g'$ up to a multiplication
factor which can be absorbed in the $a_p$. The parameters $A$ and $m_{3/2}$
are related to the potential $g$: $m_{3/2}\approx\kap^2\vev{g}$.
For the Polonyi type potential $m_{3/2}\approx\kap m^2$.
It is now easy to minimize the potential \eq{veffuone}. For simplicity we shall
assume that $\vev{Z_2}=0$. Then
\barr
V&=& \kap^{12} \left| 16 |Z_3|^6|Z_1|^{10}+25|Z_3|^8|Z_1|^8
\right| + m_{3/2}^2 \left| |Z_1|^2+|Z_3|^2\right|\nonum \\
&+&\kap^6 A'm_{3/2} |Z_1^5Z_3^4+h.c| +\frac{{\tilde g}^{'2}}{8}
\left( |Z_1|^2-\frac{5}{4}|Z_3|^2\right)
\earr
minimizing with respect to $Z_1$ and $Z_3$ one finds that
the equations imply
\beq
|Z_1|^2-\frac{5}{4} |Z_3|^2={\cal O}(m_{3/2}^2)
\eeq
and $\vev{Z_1}$ satisfies the equation
\beq
x^2+\left( \frac{5A'}{576}\right) m_{3/2} x+\half\left( \frac{5}{32}
\right)^2 m_{3/2}^2=0
\eeq
where $x=\kap^6\vev{Z_1}^7$. The solution of this quadratic equation gives
\barr
Z_1&\approx& {\cal O}\left( \frac{m_{3/2}}{100\,\kap^6}\right)^{\frac{1}{7}}\\
&&\simeq 10^{16}\gev\equiv M_X
\earr
{}From the above analysis we deduce that for the solution with $Z_1,Z_2$
symmetric
this gives
\barr
\vev{Z_1}^2=\vev{Z_2}^2&=&\frac{5}{8}|Z_3|^2+{\cal O}(m_{3/2}^2)\\
&=&{\cal O}(M^2_X)
\earr
so the mass for the $U(1)$ gauge bosons is ${\cal O}(M_X)$.
The approximation we used to obtain this solution is valid because shifts
in the low-energy sector due to the breaking of the $U(1)_X$
symmetry are of order
\beq
\kap^2g'\approx{\cal O}(\kap^8M^9_X) = {\cal O}(1\gev)
\eeq

For the solution \eq{bl} there is no superpotential and the above mechanism
will not apply. It is still possible to break supersymmetry and $U(1)_X$
by taking a non-minimal kinetic energy of the form
\beq
\sum_{i=1}^3(|Z_i|^2+\alpha|Z_i|^4)-\ln\sum_{i=1}^3|Z_i|^2.
\eeq
However, this would only lead to solutions
$\vev{Z_1}=\vev{Z_2}=\vev{Z_3}={\cal O}(\frac{1}{\kap})$. As will be clear
from the discussion in the following section these solutions will lead to
terms in the effective potential which give rise to too fast proton
decay since $M_X\kap={\cal O}(1)$. It does not seem that one can construct a
realistic model in this case.

Since the \ux-symmetry is broken, there will be \ux-breaking terms
in the effective action, and these will be induced through the \ux-gauge and
gaugino interactions. Such terms can lead to proton decay, or to baryon and
lepton number non-conserving processes. We discuss them in the next section.

\section{Baryon- and Lepton-Number Violation}
\label{sec:blviol}
The baryon- and lepton-number violating terms of
Eqs.\eqs{dim4}{dim5} and can lead to proton decay. We first discuss the
dimension four terms \eq{dim4}. Given our solutions from Section
\ref{sec:cancel} their charges $Q_X$ can be expressed in terms of $l,e$,
\beq
Q_X(LL\Ebar ,LQ\Dbar,\Ubar\Dbar\Dbar,L\Hbar)= 2l+e.
\eeq
For a non-trivial $U(1)_X$ we had required $2l+e\not=0$ and thus at tree-level
all dimension four terms are excluded. However, local supersymmetry is a
non-renormalizable theory. We thus expect our model to be the effective action
of a more complete unified model at a higher scale. We expect this unified
theory to include $\Delta B,\Delta L\not=0$ effects. The non-renormalizable
terms  in our model below the Planck scale will then be obtained as the 1-loop
effective action of this unified theory. However, for lack of knowledge of
the unified model we shall consider the most general non-renormalizable
interactions between the singlets $Z_i$ and the MSSM
fields $S_i$ which are $G_{SM}\otimes U(1)_X$ gauge-invariant
\beq
\kap^{N}Z_1^{n_1} Z_2^{n_2}Z_3^{n_3}  \prod_{i=1}^{3}S_i
\lab{nonrenorm}
\eeq
where $N=n_1+n_2+n_3$ and the $S_i$ product is a trilinear term of
Eq.\eq{dim4}.\footnote{Since the low-energy effective Lagrangian
always contains the term $H\Hbar$, and the terms $L\Hbar,LL\Ebar,LQ\Dbar$
are either present or absent simultaneously, it is possible in our models
to rotate away the $L\Hbar$ terms and we do not further consider them.
However, we emphasize that for other models this must always be checked in
each case.} After breaking $U(1)_X$ in principle one has to
find the one-loop effective action below $M$, which will include
$\Delta B,\Delta L\not=0$ terms. As our singlet interactions are
non-renormalizable from the start, a calculation of the effective
interaction is not very reliable. But all terms in the effective action are
necessarily derivable from non-renormalizable terms of the form
\eq{nonrenorm}. We shall then consider the terms obtained from \eq{nonrenorm}
by replacing $Z_i\ra\vev{Z_i}$. If all the $\Delta B,\Delta L\not=0$ effects
generated in this manner are suppressed, then we can conclude that $\Delta
B,\Delta L\not=0$ effects are not observable in the laboratory.

When replacing $Z_i\ra\vev{Z_i}$ we obtain an effective superpotential
containing the terms
\beq
(\kap M_X)^N \prod_{i=1}^{3}S_i
\eeq
with effective coupling constants $\eps^N=(\kap M_X)^N \ll1$. Here
$M_X\approx\vev{Z_i}
\approx 10^{16}\gev$ is the scale of $U(1)_X$ breaking. Since the terms
$LQ\Dbar$ and $\Ubar\Dbar\Dbar$ have the same charge $Q_X$ they will be
suppressed by the same power of $\lam\sim\eps^N$. If we have an effective
superpotential
\beq
g^{eff}(S_i)
=\lam LL\Ebar+\lam' LQ\Dbar+\lam''\Ubar\Dbar\Dbar
\lab{rparity}
\eeq
then we can use previous proton decay rate calculations \cite{langacker}
and the experimental lower bound on the proton lifetime \cite{pdg} to
estimate the extremely strict bound
\beq
\lam\cdot \lam' < 10^{-27} \left(\frac{m_{squark}}{200\gev}\right)^2.
\eeq
Since $\eps\approx 10^{-3}$ we must have $N\geq5$.
In our first model $(z_1,z_2,z_3)=(1,-\frac{4}{5},-\frac{4}{5})$,
$2l+e=\frac{1}{5}$ and the lowest dimensional term is
\beq
\kap^7 Z_1^3(Z_2Z_3)^2 \,\, g^{eff}(S_i),
\eeq
and indeed we get the appropriate suppression with $N=7$. For our second model,
$(z_1,z_2,z_3)=(1,1,-\frac{5}{4})$, $2l+e=-\frac{1}{4}$ and the lowest
dimensional term is
\beq
\kap^7 Z_1^3 Z_2^4 \,\,g^{eff}(S_i),
\eeq
and again $N=7$. The general case for rational $q$ is not soluable. However,
we can determine an upper bound on the suppression. For terms symmetric in
$Z_2,Z_3$
we obtain
\beq
\kap^{2n_1+n_3} (Z_1Z_2)^{n_1} Z_3^{n_3}\,\,g^{eff}(S_i)
\eeq
which has the charge
\beq
Q=\frac{-8n_1}{5+3q^2}+n_3+\frac{1-q^2}{5+3q^2}\equiv0
\eeq
and must vanish. For $q\not=1$ the lowest dimensional solutions are $n_1=3,n_3
=1$ and $n_1=2,n_3=3$, and thus $N=7$ once again. However, it is possible that
for certain values of $q$ the suppression might be weaker.

We now turn to the dimension five terms. Their charges are given by
\barr
Q_X(LH\Hbar\Hbar,\Hbar\Hbar\Ebar^*)=
\half Q_X(LL\Hbar\Hbar)&=&2l+e
\lab{dim5safe}\\
Q_X(QQQH,HQ\Ubar\Ebar,\Hbar^*,Q\Ubar L^*,\Ubar\Dbar^*\Ebar)
&=&-(2l+e) \lab{dim5safe2}\\
Q_X(QQQL,\Ubar\Ubar\Dbar\Ebar)&=&0 \lab{dim5dan}
\earr
For the terms in Eq\eqs{dim5safe}{dim5safe2} the discussion is identical
to that of the dimension four terms above. They are suppressed by $\eps^7$ in
any effective theory. The last two terms are $B-L$ invariant and
they can not be excluded by any gauged $U(1)_X$, given our assumptions.
It is however not so clear whether these terms actually pose a problem. In
Ref.\cite{weinberg} the partial proton decay rate via the operator $(QQQL)_F$
was estimated to be
\beq
\Gamma_p(QQQL)\approx \left( \frac{k^2 e^2}{8\pi^2}\right)^2
\frac{m_{proton}^5}{(M {\tilde m})^2}.\lab{susyprot}
\eeq
Here $M$ is the mass suppression of the non-renormalizable dimension-5
term in the effective Lagrangian. For us $M=\kappa^{-1}$. $e$ is the
electric coupling which enters when the final state s-fermions are converted
via electroweak gaugino exchange to their R-parity even partners. This
involves a 1-loop diagram, hence the $8\pi^2$. ${\tilde m}$
is an effective supersymmetric mass, \eg ${\tilde m}=m_{gaugino}/m_{squark}^2$.
$k^2$ is the coupling constant of the effective operator; it is squared since
when generated from renormalizable interactions it involves at least two
coupling constants. The dominant decay mode corresponding to \eq{susyprot}
is $p\ra K^+\nu$ \cite{drw}; the experimental bound is \cite{pdg}
$\tau_{proton}>10^{32} a$. This then corresponds to
\beq
k<10^{-4} \left(\frac{{\tilde m}}{100\gev}\right)^{1/2}.
\lab{kbound}
\eeq
This is the same order as the muon Yukawa coupling in the Standard Model,
which is not explained but is generally also not considered to be unnatural.
None the less, we proceed to consider the conditions where these terms are
suppressed. In order for the terms \eq{dim5dan} to be induced at the loop level
when $U(1)_X$ is broken, they must occur through the exchange of $Z_i$
superfields. This implies these terms are of the form
\beq
f(Z_1,Z_2,Z_3)(\alpha QQQL+\beta \Ubar\Ubar\Dbar\Ebar)
\eeq
$\alpha,\beta$ are constants and $f(Z_1,Z_2,Z_3)$ must have $U(1)_X$ charge
zero. The lowest order contribution is for $f(Z_1,Z_2,Z_3)$ to be of the
same form as $g'$ in \eq{gprime}
\beq
f(Z_1,Z_2,Z_3)=\kap^{10}Z_3^4\sum_{p=0}^5 Z_1^{5-p}Z_2^p
\eeq
and this gives rise to the effective coupling
\beq
G_5=\vev{f}\sim\kap^{10}\vev{Z_i}^9\sim10^{-36}\gev^{-1}
\eeq
which is highly suppressed. A constant term corresponds to a tree-level term
and is not allowed. Thus the dimension five terms are safe as well.
However, when the unknown unified theory is broken, and before $U(1)_X$
is broken the $QQQL$ term could be generated directly in the effective
theory. In this case the coupling must satisfy the above bound \eq{kbound}.

\section{LSP Lifetime and Neutrino Masses}
We now turn to two applications of our analysis. We have found that
in general we expect an effective $R$-parity breaking superpotential
\eq{rparity} with highly suppressed but non-zero couplings
\beq
\lam\approx\lam'\approx\lam''\approx10^{-21}.
\eeq
The terms of $g^{eff}(S_i)$ induce the decay of the lightest supersymmtric
particle (LSP).\footnote{The allowed dimension 5 terms of \eq{dim5dan} conserve
R-parity and are thus irrelevant for the decay of the LSP.} For a photino
LSP the decay rate has been calculated by Dawson \cite{dawson}
\beq
\Gamma_{\phot}=\frac{\alpha\lam^2}{128\pi^2}\frac{m_{\phot}^5}{M_{{\tilde
f}}^4}.
\eeq
$\alpha$ is the fine structure constant, $\lam$ is the coupling in
$g^{eff}$ and $M_{{\tilde f}}$ is for example the scalar electron mass.
The more general decay of a neutralino LSP is given in \cite{morawitz}.
The lifetime of the LSP is then given in natural units as
\beq
\tau_{\phot}=10^{23}s \,\left(\frac{10^{-21}}{\lam}\right)^2
\left(\frac{50\gev}{m_{\phot}}\right)^5
\left(\frac{M_{{\tilde f}}}{150\gev}\right)^4.
\eeq
The large difference in lifetime compared to the proton is due to
the different masses and the two powers less dependence on $\lam$
\beq
\frac{\tau_{\phot}}{\tau_{proton}}\approx \left(\frac{m_{lsp}}{m_{proton}}
\right)^5 \left(\frac{e}{\lam}\right)^2\approx 10^{49}
\eeq
for $m_{\phot}=50\gev$; $e$ is the electric charge quantum.

If we allow a supersymmetric mass range up to $1\tev$ for the scalar
fermion mass and LSP masses up to $M_{\tilde f}$ we obtain in terms
of the lifetime of the universe $\tau_{u}=2\cdot10^{17}s$
\beq
10^{3}<\frac{\tau_{\phot}}{\tau_{u}}<  10^{7}.
\lab{lifetimebound}
\eeq
There are several large uncertainties in this result. First, we have taken
the singlet vev $\vev{Z}={\cal O}(10^{16}\gev)$. The insertion of a simple
factor of three either way changes $\lam=\eps^7$ by a factor $10^{\pm3}$ and
$\tau_{\phot}$ by a factor $10^{\pm6}$! Second, we have only considered
a photino LSP. As discussed in Ref.\cite{morawitz} in detail the LSP lifetime
can vary by many orders of magnitude for a general neutralino LSP, depending
on the MSSM parameters.

Surprisingly enough there are severe constraints on a long-lived but
unstable LSP \cite{sarkarellis,gelminisarkar}, even for
$\tau_{\phot}>\tau_u$. The relic density $\Omega_{\phot}h^2$ of a stable LSP
is typically in the range $10^{-3}-10$ \cite{bottino}; if constrained to
the {\it minimal} $N=1$ local supersymmetric model it is $10^{-2}-10$
\cite{dreesnojiri}. For $\tau_{\phot}$ as large as $10^7\tau_u$ only
one in $10^7$ LSP will have decayed today. However, for such large
relic densities this is still a large number and the decay products
can lead to observable effects. In particular final-state decay neutrinos can
be
observed in the large underground detectors. For a LSP mass below
$1\tev$ the strictest bound \cite{sarkarellis} comes from the experimental
upper limit on upward-going muons from the IMB detector \cite{imb}.
In terms of the LSP branching ratio $B_{\nu_\mu}$ to muon neutrinos
this bound can be expressed as \cite{sarkarellis}
\beq
\Omega_{\phot} h^2< 4\cdot 10^{-9}{{B}_{\nu_\mu}}^{-1}
\frac{\tau_{\phot}}{\tau_u} \ln^{-1}(1+\frac{m_{\phot}}{1.5\tev})
\eeq
Assuming all operators in \eq{rparity} have equal strength the
muon-neutrino branching fraction is of order $1/5$. The lifetime
is reduced by about a factor of $35$, assuming $m_{\tilde\gamma}
<m_{top}$.\footnote{For $m_{\tilde\gamma}>m_{top}$ these numbers change
to $1/6$ and $45$; the product only changes from 7 to 7.5. }
For an LSP mass of $150\gev$ we thus obtain
\beq
\Omega_{\phot} h^2< 6\cdot 10^{-9} \frac{\tau_{\phot}}{\tau_u},
\eeq
and we see that only the upper range in Eq.\eq{lifetimebound} is
still allowed. This analysis is a bit crude in many respects. For
example, both the relic denisty and the LSP lifetime depend strongly
on the MSSM parameters and it would be appropriate to perform a
correlated analysis. This is beyond the scope of this paper.

Thus although the LSP lifetime is many orders of magnitude shorter than the
experimental bound on the proton lifetime ($\tau_{proton}>10^{22}\tau_u$)
the resulting bound on the couplings $\lam,\lam',\lam''$ of \eq{rparity} is
actually $2-3$ orders of magnitude stricter, if all operators are of equal
strength.

\medskip

The non-renormalizable interactions \eq{nonrenorm} can also give
rise to neutrino masses. The singlets $Z_i$ can couple to the
left-handed neutrinos via
\beq
\kap^6L\Hbar \prod_i Z_i
\eeq
leading to Dirac neutrino masses of order $\eps^6\vev{\Hbar}$.
The leading Majorana neutrino mass for the singlets is given in
(fermionic) analogy to Eq.\eq{veffuone}; it is of order $m_{3/2}={\cal
O}(100\gev)$. We then obtain the see-saw mass matrix
\beq
\left(
\begin{array}{cc}
0&\eps^6\vev{\Hbar}\\
\eps^6\vev{\Hbar} & m_{3/2}
\end{array}
\right).
\eeq
The light-neutrino mass is given by
\beq
m_\nu\approx \frac{(\eps^6 \vev{\Hbar})^2}{m_{3/2}}
\approx 3\cdot 10^{-18}\,eV,
\eeq
which is extremely small and unobservable. This is consistent with all
experiments but can not explain the hints for neutrino masses such as
the solar neutrino puzzle.

\section{Conclusion}
We have found the most general anomaly-free $U(1)_X$ with three $G_{SM}$
singlets. In local supersymmetry the breaking of $U(1)_X$ is predicted to
be at the scale $10^{16}\gev$. The most general gauge-invariant
non-renormalizable superpotential is consistent with proton decay
experiments. It is only barely consistent with the cosmological
bounds on the decay of the lightest supersymmetric particle. This poses
a severe constraint for all supersymmetric models.

\bigskip

{\bf Acknowledgements}\newline
We thank Hans-Peter Nilles for several helpful discussions. We thank Subir
Sarkar for a discussion of the bounds on a long-lived but unstable LSP.

\end{document}